\documentclass[10pt, twocolumn]{article}

\usepackage{simpleConference}
\usepackage{graphicx}
\usepackage[cmex10]{amsmath}
\usepackage{amsfonts}
\usepackage{url}

\usepackage{color}

\usepackage{siunitx}
\usepackage{paralist}
\usepackage{colortbl}

\usepackage{enumitem}

\newif\ifcomments\commentstrue

\ifcomments
\newcommand{\authornotation}[3]{\textcolor{#1}{[#3 ---#2]}}
\newcommand{\todo}[1]{\textcolor{red}{[TODO: #1]}}
\else
\newcommand{\authornotation}[3]{}
\newcommand{\todo}[1]{}
\fi

\newcommand{\wss}[1]{\authornotation{blue}{SS}{#1}}

\newcommand{\progname}{SFS}
\newcommand{\colAwidth}{0.16\textwidth}
\newcommand{\colBwidth}{0.89\textwidth}

\begin{document}

\title{Debunking the Myth that Upfront Requirements are Infeasible for
  Scientific Computing Software}

\author{Spencer Smith and Malavika Srinivasan,
Computing and Software Department,\\
McMaster University, Canada\\
Email: smiths@mcmaster.ca, sriniva@mcmaster.ca
\and
Sumanth Shankar,
Mechanical Engineering Department,\\
McMaster University, Canada\\
Email: shankar@mcmaster.ca }

\maketitle
\begin{abstract}

  Many in the Scientific Computing Software community believe that upfront
  requirements are impossible, or at least infeasible.  This paper shows
  requirements are feasible with the following:
\begin{inparaenum}[i)]
\item an appropriate perspective (`faking' the final
  documentation as if requirements were correct and complete from the start, and
  gathering requirements as if for a family of programs);
\item the aid of the right principles (abstraction, separation of concerns,
  anticipation of change, and generality);
\item employing SCS specific templates (for Software Requirements and Module
  Interface Specification); 
\item using a design process that enables change (information hiding); and,
\item the aid of modern tools (version control, issue tracking, checking,
  generation and automation tools).
\end{inparaenum}
Not only are upfront requirements feasible, they provide significant benefits,
including facilitating communication, early identification of errors, better
design decisions and enabling replicability.  The topics listed above are
explained, justified and illustrated via an example of software developed by a
small team of software and mechanical engineers for modelling the solidification
of a metal alloy. 
\end{abstract}

\noindent \textbf{keywords: }
software engineering; scientific computing; requirements analysis;
  information hiding; documentation; casting

\let\thefootnote\relax\footnotetext{Presented at SE4Science 2019 (ICSE
    Workshop), \url{https://se4science.org/workshops/se4science19/}}

\section{Introduction} \label{SecIntro}

Upfront requirements are often considered infeasible for Scientific Computing
Software (SCS).  The truth is that requirements for SCS are no more challenging
than for any other domain -- requirements are difficult for everyone.  What is
needed is not excuses, but careful, rigorous, sometimes painful thought and
effort, as early and as often as possible.  With a change in attitude and
perspective, the right tools, templates and principles, upfront requirements are
possible for SCS.

Throughout this paper, the term upfront requirements does not literally mean
requirements documentation that is entirely complete and correct as the first
activity in the development process.  Rather, upfront requirements means that
requirements are considered early in the design process and the documentation is
`faked'~\cite{ParnasAndClements1986} as if requirements were the first step.

Requirements are documented in a Software Requirements Specification (SRS).  An
SRS describes the functionalities, expected performance, goals, context, design
constraints, external interfaces, and other quality attributes of the
software~\cite{IEEE1998}. In a scientific context, an SRS records the necessary
terminology, notations, symbol definitions, units, sign conventions, physical
system descriptions, goals, assumptions, theoretical models, data definitions,
instance models and data constraints~\cite{SmithEtAl2007}.  An SRS provides many
advantages during software development~\cite{ParnasAndClements1986}. For
instance, an SRS acts as an official statement of the system requirements for
the developers, stakeholders and the end-users, and creating the SRS allows for
earlier identification of errors and omissions.  As Boehm's data
shows~\cite{Boehm1981}, early identification of errors implies a significant
return on investment, since the cost of fixing errors increases dramatically for
later development stages.  An SRS is invaluable for verification, since the
quality of software cannot be assessed without a standard against which to judge
it. Better design decisions are facilitated by the information captured in an
SRS.

Despite these benefits, an SRS is rarely written for SCS software. As the
following quotes highlight, previous research has repeatedly shown that many in
the community believe that upfront requirements are infeasible for SCS:

\begin{itemize}

\item ``Full up-front requirement specifications are impossible: requirements
  emerge as the software and the concomitant understanding of the domain
  progress.''~\cite{SegalAndMorris2008}

\item ``Since scientific software is deeply embedded into an exploratory
  process, you never know where its development might take you.  Thus, it is
  hard to specify the requirements for this kind of software up front as
  demanded by traditional software
  processes.''~\cite{JohansonAndHasselbring2018}

\item ``The research scientists ... do not appreciate the need to articulate
  requirements fully and upfront as demanded by a staged methodology, and found
  this articulation very difficult to do.''~\cite{Segal2005}

\item ``computational scientists generally adopt an agile development approach
  because they generally do not know the requirements up
  front.''~\cite{EasterbrookAndJohns2009}

\item ``Supplying requirements upfront ran counter to the [scientists] previous
  experience of developing their own software in the
  laboratory.''~\cite{Segal2007}

\item ``None of our interviewees created an up-front formal requirements
  specification. If regulations in their field mandated a requirements document,
  they wrote it when the software was almost
  complete.''~\cite{SandersAndKelly2008}

\end{itemize} 

Why is the negative attitude toward upfront requirements, and requirements in
general, so prevalent?  Part of the reasoning may be that SCS practitioners
rarely have formal training in Software Engineering (SE)~\cite{CarverEtAl2013}.
Moreover, experiments with applying SE to SCS have often made the mistake of not
tailoring the approach to scientists.  SCS developers have historically not
documented requirements, but that does not mean that they should not.  We cannot
ignore the possibility that requirements are feasible with the right principles,
techniques and tools.  Given the benefits of requirements, investigating their
feasibility is worth further effort.

This effort is justified because of quality concerns.  Faulk et
al.~\cite{FaulkEtAl2009} observe, ``growing concern about the reliability of
scientific results based on ... software.''  Embarrassing failures have
occurred, like a retraction of derived molecular protein structures
\cite{Miller2006}, false reproduction of sonoluminescent fusion
\cite{PostAndVotta2005}, and fixing and then reintroducing the same error in a
large code base three times in 20 year~\cite{MilewiczAndRayborn2018}.  A recent
report on directions for SCS research and education states: ``While the volume
and complexity of [SCS] have grown substantially ... [SCS]
traditionally has not received the focused attention it so desperately needs
... to fulfill this key role as a cornerstone of long-term collaboration and
scientific progress'' \cite{RudeEtAl2018}.  Estimates suggest that the number of
released faults per thousand executable lines of code during a given program’s
life cycle is at best 0.1, and more likely 10 to 100 times worse
\cite{Hatton2007}.


The sections of this paper present arguments for the value and feasibility of
upfront requirements for SCS.  To lend weight to the arguments, a specific
real-world software development project is cited.  The example consists of
software to model the solidification of molten metal
alloys~\cite{Srinivasan2018}.  An overview is given in Section~\ref{SecSFS}.
The arguments in favour of upfront requirements starts with the previously
mentioned idea of `faking' a rational development process
(Section~\ref{SecRationalDesignProcess}).  Following this, the specific
principles and templates for writing an SRS tailored to the needs of SCS are
presented in Section~\ref{SecSRS}.  
For upfront requirements to be effective in an environment where those
requirements are constantly changing and evolving, the approach to design also
needs to accommodate change.  Application of the principles of design for
change, along with examples, is discussed in Section~\ref{SecDesign}.
Experience has shown that principles and techniques are not enough to lead to a
change in software development; therefore, tools that facilitate upfront
requirements are summarized in Section~\ref{SecTools}.

\section{Software for Solidification} \label{SecSFS}

The running example throughout this paper is called \progname{}: Software For
Solidification~\cite{Srinivasan2018}.  This software is used to analyze
solidification data for molten metal alloys.  \progname{} was developed by this
paper's authors as a partnership between software and mechanical engineers.

SFS is typical of SCS for two reasons:
\begin{inparaenum}
\item the motivation is from a practical scientific/engineering challenge -- in
  this case, how to reduce the number of defective cast metal parts; and,
\item the high level goal (using experimentally measured temperature history
  data to understand the effect of temperature and temperature gradients on the
  solid fraction) was clear from the outset, but the details on the appropriate
  assumptions and numerical techniques were initially uncertain.
\end{inparaenum}

\begin{figure}[h!]
  \begin{center}
    {
      \includegraphics[width=0.5\columnwidth]{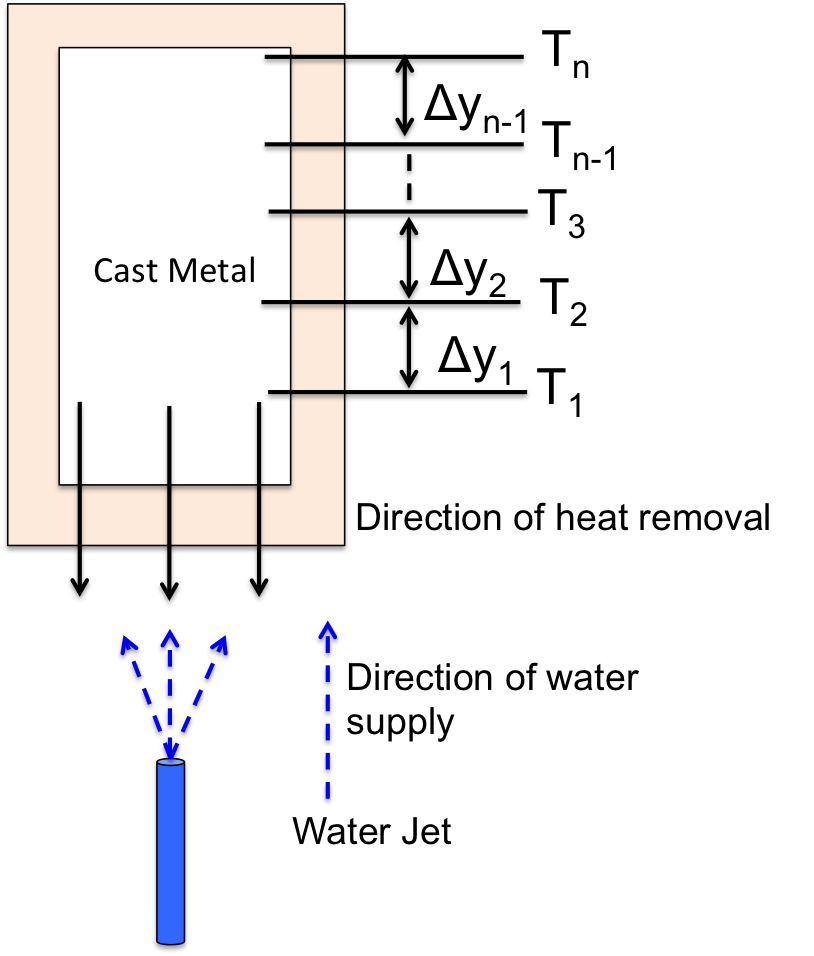}
    }
    \caption{\label{Fig_SFSDIG1} Experimental setup for cooling a liquid metal
      alloy}
  \end{center}
\end{figure}

The main input for the software consists of temperature history data at
different heights within a cylindrical sand mould containing a molten metal
alloy.  The dimensions of the cylinder are selected to ensure approximately
unidirectional heat removal.  The experimental setup is shown in
Figure~\ref{Fig_SFSDIG1}. The thermocouples for recording temperatures are
represented by the label $T_i$, where $i$ is the thermocouple number.  A water
jet is directed at the bottom of the cylinder to aid in the cooling process and
to facilitate 1D heat transfer.

\begin{figure}[h!]
	\begin{center}
		{
			\includegraphics[width=0.45\textwidth]{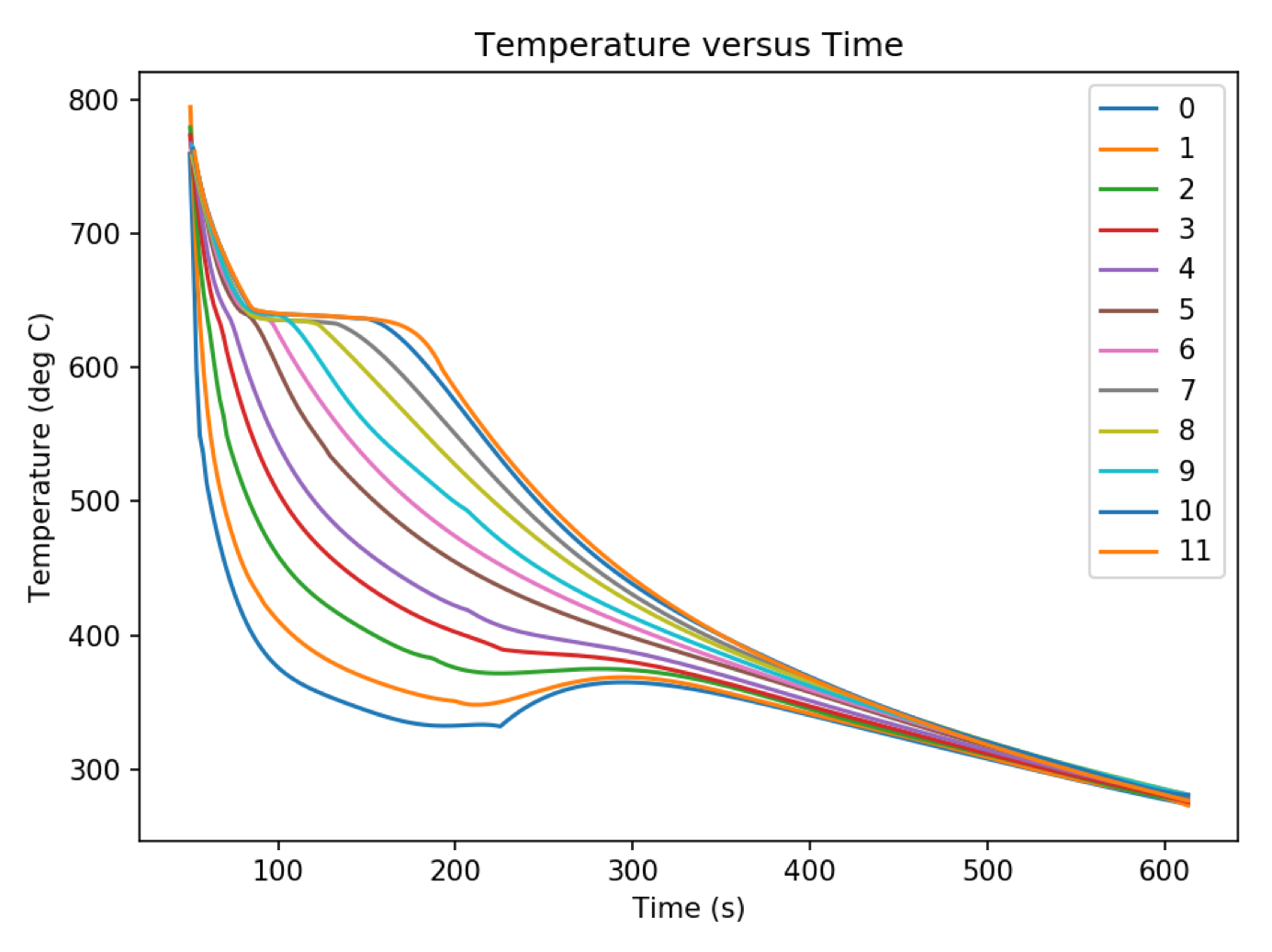}
		}
		\caption{\label{Fig_SFSDIG2} Typical cooling curves}
	\end{center}
\end{figure}

During solidification, heat is given out from the liquid phase to form a solid.
During this process, the alloy undergoes a phase transition from liquid to 2
phase (solid + liquid) and finally to solid. Typical temperature data for the
sequence of thermocouples is shown in Figure~\ref{Fig_SFSDIG2}.  The numbering
of the thermocouples increases as the distance from the bottom of the cylinder
increases.  For any instant of time, the thermocouples with the higher
temperatures and slower rates of cooling are those nearer the top of the
cylinder.

The output of the software is the fraction solid, either as a function of time,
temperature or cooling rate.  The fraction solid ranges from 0 (liquid) to 1
(solid).  The fraction solid is obtained by solving an Ordinary Differential
Equation (ODE) based on the temperature data, material properties and processing
conditions~\cite{Srinivasan2018}.




	 
	 


\section{Faking a Rational Design Process} \label{SecRationalDesignProcess}

`Upfront requirements' are (almost always) impossible, \emph{if} the term means
software development has to start with a complete and correct statement of
requirements.  However, `upfront requirements' are possible, if the definition
is softened to mean requirements documentation is started as early as possible,
and the documentation is continuously maintained, as if a rational design
process has been followed~\cite{ParnasAndClements1986}.  The almost
impossibility of having perfect information at the start of a project does not
justify not writing requirements.  Eventually we will understand the problem at
hand -- the documentation should be (re)written from that perspective.

An argument against documentation is that developers consider reports for each
stage of software development as counterproductive~\cite[p.~373]{Roache1998}.
However, reports are only counterproductive if the development process has to
follow the same rational process implied by the documentation.  In SCS (and
other software domains for that matter) frequent change is inevitable.  As
Parnas and Clements~\cite{ParnasAndClements1986} point out, the most logical way
to present the documentation is still to `fake' a rational design process.
``Software manufacturers can define their own internal process as long as they
can effectively map their products onto the ones that the much simpler, faked
process requires''~\cite{MaibaumAndWassyng2008}.  Reusability and
maintainability are important qualities for scientific software.  Documentation
that follows a faked rationale design process is easier to maintain and reuse
because the documentation is understandable and standardized.  Understandability
is improved because the faked documentation only includes the `best' version of
any artifacts, with no need to incorporate confusing details around the history
of their discovery~\cite{ParnasAndClements1986}.  Standardization on any
process, with a rational process being a logical choice as a standard,
facilitates design reviews, change management and the transfer (and
modification) of ideas and software between
projects~\cite{ParnasAndClements1986}.

\begin{figure}[h!]
  \begin{center}
    {
      \includegraphics[width=1.0\columnwidth]{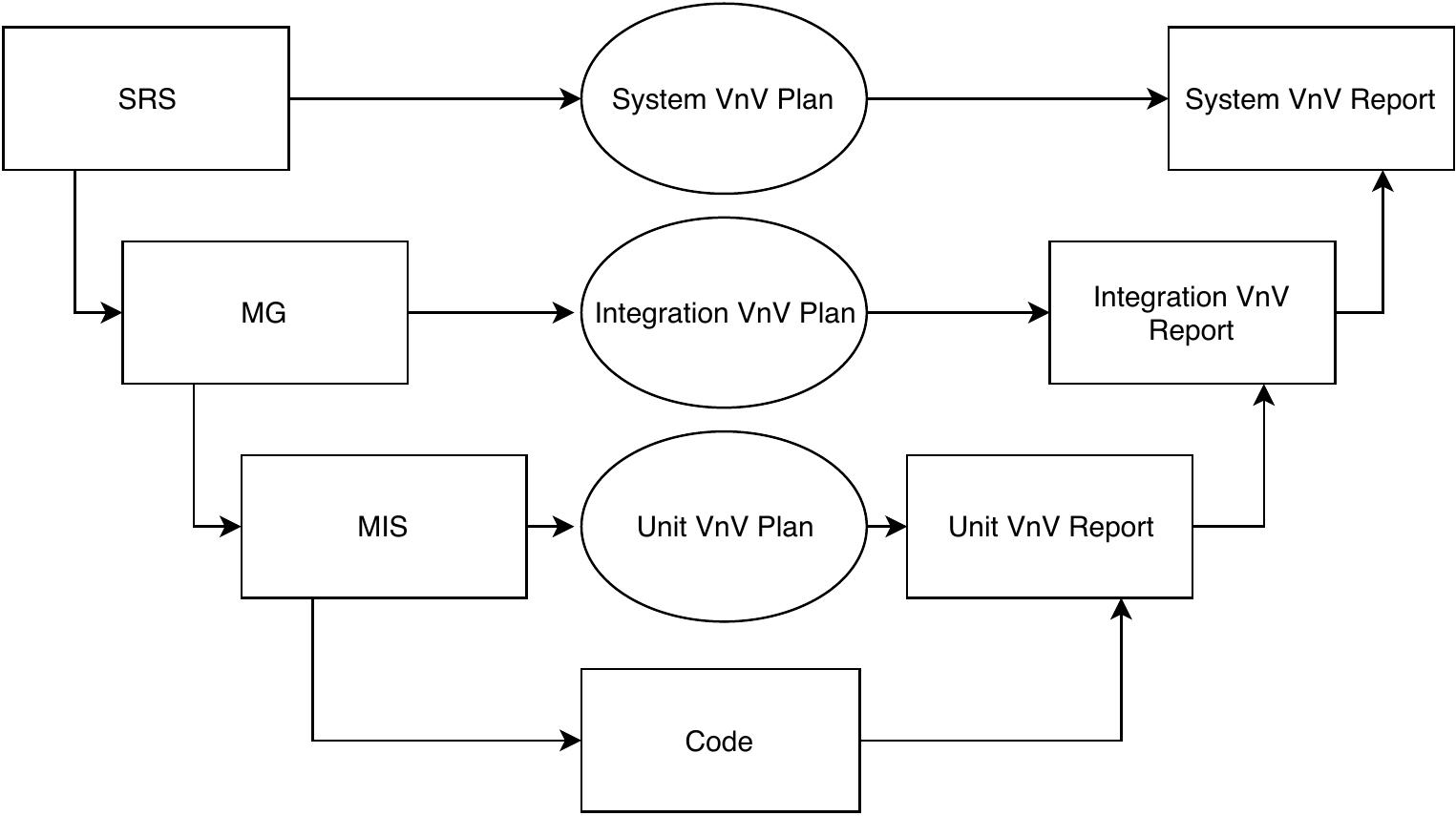}
    }
    \caption{\label{Fig_VModel} Rational design process, following the V-Model}
  \end{center}
\end{figure}

The rational design process promoted here (Figure~\ref{Fig_VModel}), and
employed for \progname{}, is a variation on the waterfall model described by
Parnas and Clements~\cite{ParnasAndClements1986}.  This V-model variation has a
testing related phase associated with each step in the typical waterfall
process.  The steps include writing requirements in a Software Requirements
Specification (SRS), a software architecture in the Module Guide (MG), and the
detailed design in the Module Interface Specification (MIS)~\cite{Smith2016,
  SmithAndYu2009}.  Each of these steps has a corresponding Verification and
Validation (VnV) plan and associated report.

Upfront requirements are possible if the documentation is maintained from this
`faking it' perspective.  However, the documentation will only be feasible if
the requirements (Section~\ref{SecSRS}) and design documentation
(Section~\ref{SecDesign}) facilitate change.  Tool support
(Section~\ref{SecTools}) will also be critical, or the frequent changes will be
overwhelming.  For the SFS example, we started with the SRS, then proceeded to
the code, and then refined the requirements and began the design documentation.
Following this first ``pass,'' the focus for subsequent development switched
frequently between documents, code and testing.  However, the documentation was
always maintained as if all of the steps occurred in a rational order.

\section{Requirements for Change} \label{SecSRS}

Upfront requirements can be feasible with a template tailored to the needs of
SCS (Section~\ref{SecTemplate}) and the use of the following principles:
Abstraction (Section~\ref{SecAbstraction}), Separation of Concerns
(Section~\ref{SecSepOfConcerns}), Anticipation of Change
(Section~\ref{SecAnticipOfChange}), Generality (Section~\ref{SecGenerality}) and
Replicability (Section~\ref{SecReplicability}).  The first four principles come
from SE~\cite{GhezziEtAl2003}, while the last is the cornerstone of the
scientific method.

\subsection{SRS Template} \label{SecTemplate}

Writing an SRS generally starts with a template, which provides guidelines and
rules for documenting the requirements.  Several existing templates contain
suggestions on how to avoid complications and how to achieve qualities such as
verifiability, maintainability and reusability~\cite{IEEE1998, ESA1991,
  NASA1989}.  Templates are generally selected to suit the problem domain.  A
template designed for the needs of SCS~\cite{SmithEtAl2007, SmithAndLai2005} was
selected for documenting \progname{}, as illustrated in Figure~\ref{SRS_Dig}.
The recommended template is suitable for science, because of its hierarchical
structure, which decomposes abstract goals to concrete instance models, with the
support of data definitions, assumptions and terminology.  Excerpts from the SRS
for \progname{} will be used in the sections below to illustrate how the
template and the writing of the SRS supports the SE and scientific principles
that make upfront requirements documentation feasible.

\begin{figure}[pt]
	
  \scalebox{1.0}{ \fbox{
      \begin{minipage}{0.95\columnwidth}
        \begin{enumerate}[leftmargin=*]
					
        \item{Reference Material:}
          \begin{inparaenum}[a\upshape)]
          \item{Table of Units}
          \item{Table of Symbols}
          \item{Abbreviations and Acronyms}
          \end{inparaenum}
					
        \item{Introduction:}
          \begin{inparaenum}[a\upshape)]
          \item{Purpose of Document}
          \item{Scope of Requirements}
          \item{Intended Audience}
          \item{Organization of Document}
          \end{inparaenum}
				
        \item{Background}
					
        \item{General System Description:}
          \begin{inparaenum}[a\upshape)]
          \item{System Context}
          \item{User Characteristics}
          \item{System Constraints}
          \end{inparaenum}
					
        \item {Specific System Description:}\\
          \begin {inparaenum}[a\upshape)]
          \item Problem Description:
            \begin {inparaenum}[i\upshape)]
            \item Terminology and Definitions
            \item Physical System Description
            \item Goal Statements
            \end{inparaenum}
					
          \item {Solution Characteristics Specification:}
            \begin {inparaenum}[i\upshape)]
            \item Assumptions
            \item Theoretical Models
            \item General Definitions
            \item Data Definitions
            \item Instance Models
            \item Data Constraints
            \item Properties of a Correct Solution
            \end{inparaenum}
          \end{inparaenum}
				
        \item{Requirements:}\\
          \begin{inparaenum}[a\upshape)]
          \item {Functional Requirements:}
            \begin {inparaenum}[i\upshape)]
            \item Configuration Mode
            \item Calibration Mode
            \item Calculation Mode\\
            \end{inparaenum}
          \item {Non-Functional Requirements:}
            \begin {inparaenum}[i\upshape)]
            \item Look and Feel Requirements
            \item Usability and Humanity Requirements
            \item Installability Requirements
            \item Performance Requirements
            \item Operating and Environmental Requirements
            \item Maintainability and Support Requirements
            \item Security Requirements
            \item Cultural Requirements
            \item Compliance Requirements
            \end{inparaenum}
          \end{inparaenum}
				
        \item{Likely Changes}
				
        \item {Unlikely Changes}
				
        \item{Supporting Information}
				
        \end{enumerate}
      \end{minipage}
    } }
  \caption{Table of Contents for SRS}\label{SRS_Dig}
\end{figure}

\subsection{Abstraction} \label{SecAbstraction}

Abstraction focuses on what is important, while ignoring what is
irrelevant~\cite{GhezziEtAl2003}.  Requirements at the right abstraction level
lead to a model where the likely changes are excluded as irrelevant details.
With the aid of the principle of abstraction, the SRS for \progname{} underwent
very little modification over the life of the project.  Although considerable
exploration and experimentation were necessary, this exploration took place
through the design and implementation, not the SRS.

The overall goal for \progname{} (Figure~\ref{SRS_Goal}) was written so that it
would remain true throughout the project.  The goal statement does not list
specific material properties or initial conditions, since they may change as
different models are explored.

\begin{figure}[pt]	
  \scalebox{1.0}{ \fbox{
      \begin{minipage}{0.95\columnwidth}

        \noindent For a given experiment with a metal alloy, using the
        thermocouple locations, temperature readings, material properties and
        initial conditions, \progname{}:
        \begin{itemize}
        \item[GS1:] Computes the solid fraction ($f_s$) as a function of
          temperature ($T$) and cooling rate ($f_s(T, \frac{dT}{dt})$).
        \end{itemize}

      \end{minipage}
    } }
  \caption{Goal Statement}\label{SRS_Goal}
\end{figure}

The goal statement mentions temperature readings.  This concept needs to be
further refined to how discrete readings are transformed into a continuous
approximation of the evolution of the temperature field over time.  This is done
through the general definition (GD4) of the transformation of the experimental
data (Figure~\ref{Fig_Transform}).  GD4 abstracts the concept of taking
experimental data ($\mathit{Tdata}$) at the thermocouples (shown in
Figure~\ref{Fig_SFSDIG2}) over time and determining a function $T(y, t)$.  This
function takes the position $y$ (\si{\metre}), as measured from the bottom of
the cylinder, and the time $t$ (\si{second}) and returns the temperature
(\si{\celsius}).  $\text{fit}()$ is a function that takes the thermocouple data
$\mathit{Tdata}$, the locations of the thermocouples $y_{TC}$ and the time step
$dt$, and returns the appropriate function $T(y, t)$. $m$ is the number of
instants of time where the thermocouple data is measured and $n$ is the number
of thermocouples.  The specific approach for fitting, such as interpolation
versus regression, is not specified, since these details are not relevant at
this stage.  All that needs to be known is that there will be a function for
finding the temperature at any point in space and time.

\begin{figure}[pt]
  \scalebox{1.0}{
    \begin{minipage}{0.86\columnwidth}

\renewcommand*{\arraystretch}{1.5}
\begin{tabular}{| p{\colAwidth} | p{\colBwidth}|}
\hline
\rowcolor[gray]{0.9}
Number& GD4\\
\hline
Equation&$ T(y, t) = \text{fit}(\mathit{Tdata}, y_{TC}, dt)$ where\newline
          $\text{fit}: \mathbb{R}^{m \times n} \rightarrow \mathbb{R}^n
          \rightarrow \mathbb{R} \rightarrow (\mathbb{R}
          \rightarrow \mathbb{R} \rightarrow \mathbb{R}) $\\
\hline
Descript. &
...\\
\hline
\end{tabular}

    \end{minipage}
  }
  \caption{General definition of transforming experimental
    data}\label{Fig_Transform}
\end{figure}

The function $T(y, t)$ is used to find the solid fraction $f_S$, as given in the
excerpt of an instance model (Figure~\ref{SRS_IM}).  The instance models are the
closest the SRS gets to code, since they list the program inputs and outputs.
Abstraction is still employed though, since the numerical algorithm for finding
partial derivatives of $T(y, t)$ is not given, nor is the algorithm for solving
the ODE.  The cross-references to Data Definitions (DD) in Figure~\ref{SRS_IM}
show the interrelation between the SRS parts.

\begin{figure}[pt]
  \scalebox{1.0}{ 
      \begin{minipage}{0.95\columnwidth}
        \renewcommand*{\arraystretch}{1.5}
        \begin{tabular}{| p{0.16\textwidth} | p{0.8\textwidth}|}
          \hline
          \rowcolor[gray]{0.9}
          Number& IM4\\
          \hline
          Input & $T(y,t)$ (see~DD4), from which $\frac{\partial
                  T}{\partial t}$ and
                  $\frac{\partial^2 T}{\partial  y^2}$ can be derived, as required\\
                & Material properties $C_v^L(T)$, $C_v^S(T)$, $\rho_L(T)$,
                  $\rho_S(T)$, 
                  $\alpha_b$ (from IM2), $\alpha_e$ (from IM3), and $L$\\
                & $y^*$, $(t_L, T_L)$ from~DD6 and $(t_S, T_S)$
                  from~DD7\\

          \hline
          Output& Solve $f_s(t)$ at location $y^*$ such that the following ODE is
                  satisfied with
                  $f_s(t_L)=0$:\newline
                  ~\newline
                  $\dot{f_s}(f_s, t) = \frac{C_v(f_s)}{L \rho(f_s)} \left [
                  \frac{\partial
                  T(t)}{\partial t} - \alpha(f_s)\frac{\partial^2 T(t)}{\partial y^2}
                  \right ] \text{where...}$\newline\\
          \hline

\end{tabular}
\end{minipage}
} 
\caption{Instance model for finding fraction solid}\label{SRS_IM}
\end{figure}

\subsection{Separation of Concerns} \label{SecSepOfConcerns}

With the principle of separation of concerns, we reduce a complex problem to a
set of simpler problems~\cite{GhezziEtAl2003}.  This principle was used in the
design of the SRS template~\cite{SmithAndLai2005, SmithEtAl2007}.  For instance,
the different sections of the table of contents (Figure~\ref{SRS_Dig}) can each be
considered separately.  As an example, thinking about functional requirements,
like the governing equations for physical models, is made easier if one
momentarily neglects consideration of nonfunctional requirements, like
performance and portability.  

Upfront requirements become much more stable if the SRS documents the physical
models, but does not address numerical methods.  The decisions on numerical
techniques, such as the ODE solver, should be postponed to the design stage.
Knowing the most appropriate numerical technique is difficult at the outset, and
the initial choice is likely to change.  The physics on the other hand, if
written with the right abstraction, is much less likely to change.

Separation of concerns allows us to improve the overall software quality and
software developer productivity because it allows us to distribute the human
effort and thought throughout the development process.  Human nature is to put
off difficult tasks.  For instance, a developer might prefer to develop code
rather than think about how to test their code.  However, many testing concerns
do not have to wait until the code is written.  The associated thought and
effort can happen upfront.  For instance, the properties of a correct solution
(Section 5.vii of the SRS) are often known at the outset of a project.  These
properties should be documented, and used as the basis for test cases in the
System VnV Plan (Figure~\ref{Fig_VModel}).  What the SRS template terms
`properties of a correct solution' is often called initial hypotheses and
metamorphic tests by others~\cite{Patrick2016}.

\progname{} provides an example of a property of a correct solution for the
temperature field $T(y, t)$, which is shown graphically in
Figure~\ref{Fig_SFSDIG2} and described in Figure~\ref{Fig_Transform}.  Physics
imposes the property that for any $t$, $t > 0$,
$(\forall\, y_1, y_2: \mathbb{R} | 0 \leq y_1 \leq H \wedge 0 \leq y_2 \leq H
\wedge y_1 < y_2: T_t(y_2) \geq T_t(y_1))$ where $T_t = \lambda y: T(y, t)$.
That is $T_t(y)$ is non-decreasing with increasing $y$ ($\lambda$ is an
anonymous function).  This property was known at the outset of the project and
provided assertions in the unit tests that later detected a mistake in the
calibration of the thermocouples.

\subsection{Anticipation of Change} \label{SecAnticipOfChange}

\begin{figure}[h!]
  \begin{center}
    {
      \includegraphics[width=1.0\columnwidth]{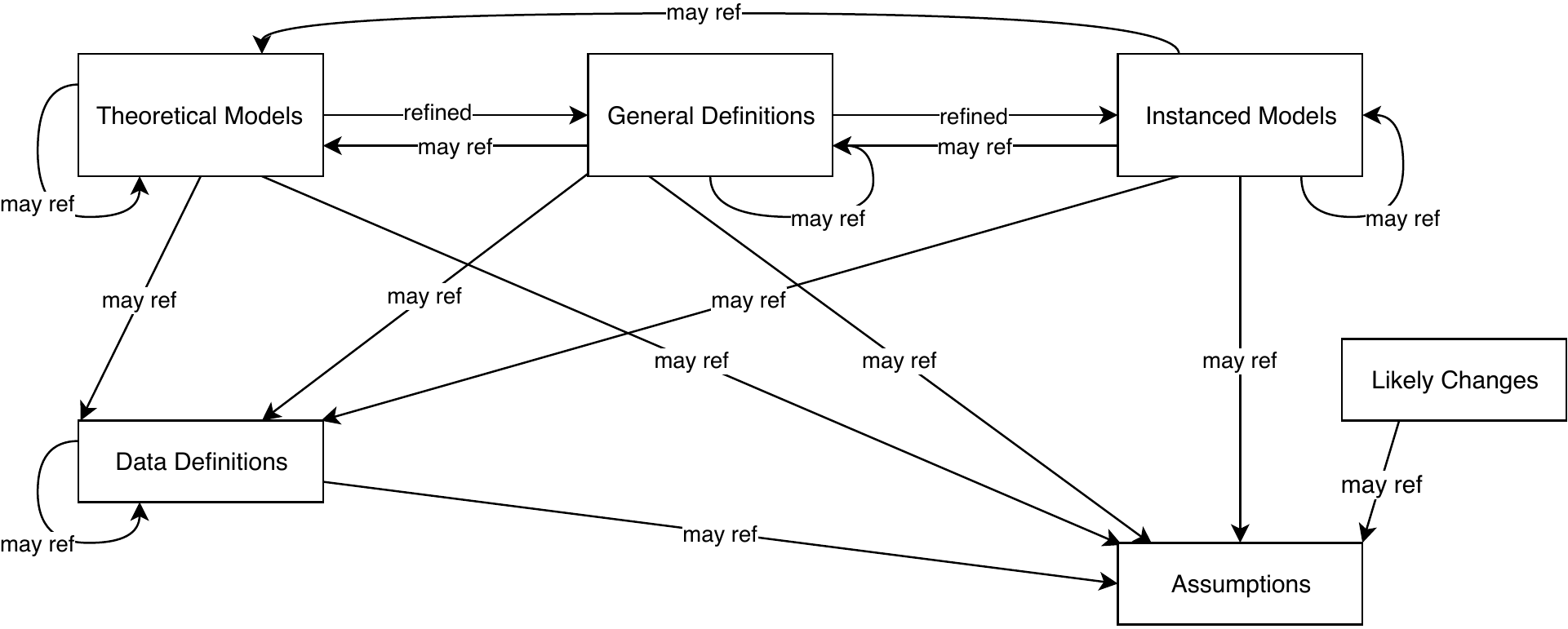}
    }
    \caption{\label{Fig_RelationsBWChunks} Relationship between parts of the SRS}
  \end{center}
\end{figure}

Upfront requirements are possible, if we anticipate future changes.  The SRS
template used for \progname{} was designed with change in
mind~\cite{SmithEtAl2007, SmithAndLai2005}.  For instance, assumptions are
listed with their traceability to other parts of the SRS explicitly indicated.
As an example for \progname{}, one of the assumptions is that the density of the
alloy can be expressed as a linear combination of the density values at the
beginning and end of the solidification.  This assumption maps to the definition
of density and the models that use density for their calculations.  If this
assumption should change, the consequences of this change are clearly indicated.
Other aspects of the template that support change include explicit traceability
between the information in the SRS, as indicated in
Figure~\ref{Fig_RelationsBWChunks} (documented by cross-references and a full
traceability matrix), a section on likely changes (for instance, the heat
transfer at the bottom of the cylinder may later be changed from constant to
temperature dependent), and the use of symbolic constants.

The template anticipates change, but successfully using it requires a new
perspective where the author(s) think of their task as documenting a family of
SCS programs, instead of just one.  As for the properties of a correct solution
mentioned in Section~\ref{SecSepOfConcerns}, this will mean intense
effort/thought at the beginning of the development process.  A program family
approach, where commonalities are reused and variabilities are identified and
systematically handled, is natural for scientific
software~\cite{SmithMcCutchanAndCao2007}.  The laws of physics are stable; they
are almost universally accepted, well understood, and slow to change.  At the
appropriate abstraction level, many problems have significant commonality, since
a large class of physical models are instances of a relatively small number of
conservation equations (conservation of energy, mass and momentum).  The
challenge is to know which simplifying assumptions are appropriate.  Therefore,
as mentioned above, the assumptions need to be documented clearly, and explicit
traceability is required to show which parts of the model they influence.  For
general purpose SCS, like a solver for a system of linear equations, the program
family should generally be clear from the start, since general purpose tools are
based on well understood mathematics.

\progname{} has benefitted from application of the principle of anticipation of
change because several sections are straightforward modifications of previous
SRS documents.  For instance, \progname{} shares verbatim the abstract
theoretical model for conservation of thermal energy with projects on modelling
a fuel pin in a nuclear reactor~\cite{SmithAndKoothoor2016} and predicting the
temperature of a solar water heating tank
(text{https://github.com/smiths/swhs}).

\subsection{Generality} \label{SecGenerality}

To aid future reuse and to facilitate software design and implementation, the
SRS should be written with generality.  This often involves using abstraction,
which is why the abstract theoretical model of conservation of thermal energy
can be reused in multiple projects (Section~\ref{SecAnticipOfChange}).

A key example of generality in \progname{} is found in the decision to write the
instance model on the calculation of $f_S$ in the standard form of a system of
ODEs (Figure~\ref{SRS_IM}).  Other work related to calculation of $f_S$ does not
target this standard form~\cite{SahinEtAl2005}.  As a consequence, this other
work requires development of an ad hoc algorithm for determining the solid
fraction.  The design and implementation of \progname{} on the other hand can,
because of its generality, take advantage of the wealth of numerical ODE solver
libraries available.

\subsection{Replicability} \label{SecReplicability}

Although reproducibility is the cornerstone of the scientific method, until
recently it has not been treated seriously in
software~\cite{BenureauAndRougier2017}.  Fortunately, in recent years multiple
conferences, workshops and individuals are calling for dramatic
change~\cite{BaileyEtAl2016}.  Reproducibility problems are even more extreme
when the goal is replicability.  A third party should be able to repeat a study
using only the description of the methodology from a published article , with no
access to the original code or computing
environment~\cite{BenureauAndRougier2017}.  However, replicability is rarely
achieved, as shown for microarray gene expression \cite{IoannidisEtAl2009} and
for economics modelling \cite{IonescuAndJansson2013}.  Replicability is not
achieved because journal papers cannot possibly document all definitions,
assumptions, models etc.  An SRS addresses completeness and ambiguity problems,
since a template is followed, there is no page limit and, SE principles are
followed.  The goal is to capture all of the required knowledge, including
derivation of equations and rationales.  If the SCS community is serious about
replicability, documentation of an SRS at the level of detail described in this
paper is required.



\section{Design for Change} \label{SecDesign}

Upfront requirements, using the principle of `faking it,' are only feasible if
the design can be modified in response to changing requirements.  This is
possible by applying the same principles listed in the previous section.  The
design of \progname{} is captured in the Module Interface Specification (MIS).
This document follows a previous template~\cite{SmithAndYu2009}.  This template
is based on a textual design specification~\cite[Chapter 4]{GhezziEtAl2003} and
the mathematical formalism of a module state
machine~\cite{HoffmanAndStrooper1995}.

An important module in \progname{} is the one responsible for calculating the
temperature at any point in space and time by using the temperature history data
at the thermocouples, as shown in the SRS under general definition GD4
(Figure~\ref{Fig_Transform}).  The design uses a special technique for
anticipation of change: information hiding~\cite{Parnas1972a}.  The changeable
secrets of the modules are hidden behind the stable interface.  This means the
design decision can be changed, without a need to modify the modules that rely
on the provided services.

Separation of concerns is used in the design to handle the complexity of fitting
over two dimensions (distance and time).  First an Abstract Data Type (FunctT)
is introduced to capture the cooling curves, which are shown for each
thermocouple in Figure~\ref{Fig_SFSDIG2}.  Second an ADT is introduced for the
contours (ContrT), which consists of a sequence of cooling curves (FunctT
objects), with each curve at a different location.  Applying the principle of
generality, the modules are not specific for temperature; they can be used for
any two dimensional data.  For instance, the same approach was used in a program
to calculate the risk of glass panes failing
(\url{https://github.com/smiths/caseStudies}).

The syntax of the access programs for FunctT is given in
Figure~\ref{Fig_FunctADT_Syntax}.  The MIS template shows application of the
principle of separation of concerns, by separating the syntax and semantics.
The interface provides a constructor (new FunctT) that takes two vectors of data
and returns a FunctT object.  Accessors include minD and maxD for finding the
extreme limits of the independent data variable.  The accessor eval returns the
value of the dependent variable ($x$) given a value for the independent variable
($y$).  The syntax of FuncT shows exceptions if the data for the independent
variable data is not in ascending order, of if the number of data points in the
two sequences do not match, or if a function evaluation is sought outside of the
domain of the given data.

\begin{figure}[pt]
  \scalebox{1.0}{
    \begin{minipage}{0.95\columnwidth} 
 
        \begin{tabular}{| p{2cm} | p{1.1cm} | l | p{2.9cm} |}
          \hline
          \textbf{Routine} & \textbf{In} & \textbf{Out} & \textbf{Exceptions}\\
          \hline
          new FunctT & $X_\text{in}$:$\mathbb{R}^n$, $Y_\text{in}$: $\mathbb{R}^n$ &
                                                                                                   FunctT
                                                             &
                                                               IVarNotAscend,\newline
                                                               SeqSizeMismatch\\
          \hline
          minD & ~ & $\mathbb{R}$ & ~\\
          \hline
          maxD & ~ & $\mathbb{R}$ & ~\\
          \hline
          eval & $x: \mathbb{R}$ & $\mathbb{R}$ & OutOfDomain\\
          \hline

        \end{tabular}

      \end{minipage}
    } 
    \caption{MIS syntax for FunctT}\label{Fig_FunctADT_Syntax}
\end{figure}

The corresponding semantics for FunctT is given in
Figure~\ref{Fig_FunctADT_Semantics}.  The MIS uses abstraction, with the
functional fit represented by a state variable with a function type
($\mathbb{R} \rightarrow \mathbb{R}$).  The specific function is hidden.  To
make it easy to change, a local function is defined for interpolation (interp).
This function can be changed to any order of interpolation, or it could be
changed to regression, or a spline.  Defining a new functional fit, just
requires redefining interp.  Following the principle of information hiding, the
interface to FunctT would remain unchanged.  This facilitates experimentation,
which was necessary for determining the best approach for \progname{}.

\begin{figure}[pt]
  \scalebox{0.9}{\fbox{
      \begin{minipage}{1.1\columnwidth}

        \textbf {State Variables}\\
        f: $\mathbb{R} \rightarrow \mathbb{R}$\\
        minX: $\mathbb{R}$\\
        maxX: $\mathbb{R}$\\

\noindent new FunctT($X_\text{in}, Y_\text{in}$):
\begin{itemize}[leftmargin=*]
\item transition:
$\mbox{minX}, \mbox{maxX}, f := X_0, X_{|X|-1}, (\lambda v:
  \mbox{interp} (X_\text{in}, Y_\text{in}, v))$
\item output: $out := \mbox{self}$
\item exception: $(\neg \mbox{isAscending}(X_\text{in}) \Rightarrow
  \mbox{IVarNotAscend} |$ $|X_\text{in}| \neq |Y_\text{in}| \Rightarrow \mbox{SeqSizeMismatch})$
\end{itemize}

...

\noindent eval($x$):
\begin{itemize}[leftmargin=*]
\item output: $out := f(x)$
\item exception: $(\neg(\mbox{minX} \leq x \leq \mbox{maxX}) \Rightarrow
  \mbox{OutOfDomain}))$ 
\end{itemize}

\textbf{Local Functions}

interp: $\mathbb{R}^n \times \mathbb{R}^n \times \mathbb{R}
\rightarrow \mathbb{R}$\\
\noindent interp($X$, $Y$, $v$)\\
$\equiv \text{interpQuad}(X_{i-1}, Y_{i-1}, X_i,
Y_i, X_{i+1}, Y_{i+1}, v) \text{ where } i = \text{index}(X, v)$
  
interpQuad: $\mathbb{R} \times \mathbb{R} \times \mathbb{R} \times \mathbb{R}
\times \mathbb{R} \times \mathbb{R} \times \mathbb{R} \rightarrow \mathbb{R}$\\
\noindent interpQuad($x_0$, $y_0$, $x_1$, $y_1$, $x_2$, $y_2$, $x$) 
$\equiv y_1 + \frac{y_2 - y_0}{x_2-x_0} (x - x_1) + \frac{y_2 - 2 y_1 + y_0}{2 (x_2-x_1)^2} (x - x_1)^2$

index: $\mathbb{R}^n \times \mathbb{R} \rightarrow \mathbb{N}$ \textit{\#
  constructor ensures seq.\ is ascending}\\
\noindent index($X$, $x$) $\equiv i \text{ such that } X_i \leq x < X_{i+1}$\\

      \end{minipage}
   } } 
  \caption{MIS semantics for FunctT}\label{Fig_FunctADT_Semantics}
\end{figure}

Figure~\ref{Fig_ContourADT} shows the syntax and semantics for the MIS for
ContourADT.  The full data set is built up by adding each successive
thermocouple's data.  The accessors, eval, dydx and d2ydx2 are used to calculate
the values of $T(y, t)$, $\frac{\partial T}{\partial t}$ and
$\frac{\partial^2 T}{\partial y^2}$, respectively.  The access program slice
returns a new FuncT that would hold the temperature values through the height
of the cylinder for a given value of the dependent variable $t$.  In this case
the slice will use the same fitting algorithm as for the cooling curves.  In the
actual MIS for \progname{} the order of the interpolating polynomial was exposed
as an input variable, so that the fit through the height could be a different
order than the fit for an individual cooling curve.

\begin{figure}[pt]
  \scalebox{1.0}{\fbox{
      \begin{minipage}{1.01\columnwidth}

      \textbf{Syntax}: \emph{Exported Access Programs}\\

      \begin{tabular}{| l | p{2.3cm} | l | l |}
        \hline
        \textbf{Routine} & \textbf{In} & \textbf{Out} & \textbf{Exceptions}\\
        \hline
        ContrT &  & ~ & ~\\
        \hline
        add & s: FunctT, $z: \mathbb{R}$ & ~ & IVarNotAscend\\
        \hline
        getC & $i: \mathbb{N}$ & ~ & InvalidIndex\\
        \hline
        eval & $x: \mathbb{R}, z: \mathbb{R}$ & ~ & OutOfDomain\\
        \hline
        dydx & $x: \mathbb{R}, z: \mathbb{R}$ & ~ & OutOfDomain\\
        \hline
        d2ydx2 & $x: \mathbb{R}, z: \mathbb{R}$ & ~ & OutOfDomain\\
        \hline
        slice & $x: \mathbb{R}$ & FunctT & ~\\
        \hline
      \end{tabular}\\

      \textbf {Semantics}: \emph{State Variables}\\
      $S$: sequence of FunctT\\
      $Z$: sequence of $\mathbb{R}$\\


      new ContrT(i):
      \begin{itemize}[leftmargin=*]
      \item transition: $S, Z := < >, <>$
      \item exception: none
      \end{itemize}

      \noindent add($s, z$):
      \begin{itemize}[leftmargin=*]
      \item transition: $S, Z := S || <s>, Z || <z>$
      \item exception:
        $exc := (|Z| > 0 \wedge z < Z_{|Z|-1} \Rightarrow
        \mbox{IVarNotAscend})$
      \end{itemize}


      \noindent eval($x, z$):
      \begin{itemize}[leftmargin=*]
      \item output: $out := \mbox{self.slice}(x).\mbox{eval}(z)$
      \item exception: none 
      \end{itemize}
      ...

      \noindent slice($x$):
      \begin{itemize}[leftmargin=*]
      \item output:
        $out := \mbox{FunctT}(Z, \langle S_0.\mbox{eval}(x), ...,
        S_{|S|-1}.\mbox{eval}(x) \rangle)$
      \item exception: None
      \end{itemize}

    \end{minipage}
   } } 
 \caption{MIS for ContrT} \label{Fig_ContourADT}
\end{figure}

\section{Tool Support} \label{SecTools}

Experience has shown that principles and templates are not enough for dramatic
change in development practices.  Although the value of documentation is
recognized, there is a sense that writing and maintaining the documentation is
too great an effort~\cite{SmithJegatheesanAndKelly2016}.  The positive influence
of tool support is observed in the CRAN (Comprehensive R Archive Network)
community, where documentation generation and consistency checking tools allow a
single SCS developer to achieve similar software quality to a team of
developers~\cite{SmithEtAl2015-SS-TR}.

SCS developers should use tools for issue tracking and version control.  Issue
tracking is considered a central quality assurance
process~\cite{BeckhausEtAl2009}.  For \progname{}, git was used for version
control and GitLab for issue tracking.  The issue tracker facilitated bringing
the team knowledge together, especially when review questions were assigned to
the domain experts.

Going forward, a knowledge-based approach for scientific software development
holds promise.  Ideally, developers should be able to create high quality
documentation without the drudgery of writing and maintaining it.  A potential
solution is to generate the documentation automatically by using Domain Specific
Languages (DSLs) over a base of scientific knowledge.  DSLs and code/document
generation provide a transformative technology for documentation, design and
verification~\cite{JohansonAndHasselbring2018, Smith2018}.  DSLs allow upfront
changes to be propagated through a project and they provide checks for
consistency and completeness.  Moreover, a generative approach removes the
maintenance nightmare of documentation duplicates and near duplicates, since
knowledge is only captured once and automatically transformed as needed.

\section{Concluding Remarks} \label{SecConclusions}

Without dramatic intervention, our collective confidence SCS is due for a
collapse.  Successfully building SCS requires communication between software
developers and experts from multiple domains. Collaboration is difficult at the
best of times, and is made worse because developers avoid the upfront effort of
thinking about the requirements and the known properties of a correct solution.
Developers tend to favour handcrafted solutions over adapting SE processes,
methods and tools \cite{FaulkEtAl2009}. Handcrafted solutions do not account for
the inevitable changes in requirements, design and implementation. The twin
challenges of changing requirements and inadequate documentation conspire to
make computational results notoriously difficult to reproduce, especially in the
case of one researcher independently replicating another's
results~\cite{Smith2018}.

The solution to quality problems is applying, adapting and developing SE
principles, tools and techniques.  However, typical processes are a barrier to
progress.  ``To break the gridlock, we must establish a degree of cooperation
and collaboration with the [SE] community that does not yet exist''
\cite{FaulkEtAl2009}.  ``There is a need to improve the transfer of existing
practices and tools from ... [SE] to [SCS]. In addition, ... there is a need for
research to specifically develop methods and tools that are tailored to the
domain'' \cite{Storer2017}.  ``Use of [SE] practices could increase the
correctness of scientific software and the efficiency of its
development.''~\cite{HeatonAndCarver2015}.  The current paper has highlighted
perspectives, SE principles, templates and tools as part of a path toward
interdisciplinary SE and SCS.

\section*{Acknowledgements}

The financial support of the Natural Sciences and Engineering Research Council
(NSERC), Automotive Partnership Grant, APC 435504-12 is gratefully acknowledged.


\end{document}